\documentclass[twocolumn,english,aps,prl,superscriptaddress]{revtex4-1}
\usepackage[latin9]{inputenc}
\usepackage{amsmath}
\usepackage{amssymb}
\usepackage{graphicx}
\usepackage{esint}

\makeatletter

\usepackage{graphicx}
\usepackage{color}
\usepackage[usenames,dvipsnames]{xcolor}
\usepackage[colorlinks=true,linkcolor=Red,citecolor=Green,linktoc=page]{hyperref}
\usepackage{multirow}
\usepackage{float}
\usepackage{flushend}
\usepackage{balance}
\usepackage[varg]{txfonts}
\usepackage{ulem}
\usepackage{fancyhdr}

\makeatother

\usepackage{babel}

\newcommand*\chem[1]{\ensuremath{\mathrm{#1}}}
\newcommand*\Tc[0]{$T_c$}
\newcommand*\hbn[0]{\textit{h}-BN}

\begin{document}
\title{Sign reversing Hall effect in atomically thin high temperature superconductors}
\author{S.\,Y.\,Frank\,Zhao}
\affiliation{Department of Physics, Harvard University,
    Cambridge, MA 02138, USA}
\author{Nicola\,Poccia}
\affiliation{Department of Physics, Harvard University, Cambridge, MA 02138, USA}
\author{Margaret\,G.\,Panetta}
\affiliation{Department of Physics, Harvard University,
	Cambridge, MA 02138, USA}
\author{Cyndia\,Yu}
\affiliation{Department of Physics, Harvard University,
    Cambridge, MA 02138, USA}
\author{Jedediah\,W.\,Johnson}
\affiliation{Department of Physics, Harvard University,
    Cambridge, MA 02138, USA}
\author{Hyobin\,Yoo}
\affiliation{Department of Physics, Harvard University,
    Cambridge, MA 02138, USA}
\author{Ruidan\,Zhong}
\affiliation{Department of Condensed Matter Physics and Materials Science, Brookhaven National Laboratory,
    Upton, NY 11973, USA}
\author{G.\,D.\,Gu}
\affiliation{Department of Condensed Matter Physics and Materials Science, Brookhaven National Laboratory,
    Upton, NY 11973, USA}
\author{Kenji\,Watanabe}
\affiliation{National Institute for Materials Science, Namiki 1-1, Tsukuba, Ibaraki 305-0044, Japan}
\author{Takashi\,Taniguchi}
\affiliation{National Institute for Materials Science, Namiki 1-1, Tsukuba, Ibaraki 305-0044, Japan}
\author{Svetlana\,V.\,Postolova}
\affiliation{Institute for Physics of Microstructures RAS, Nizhny Novgorod 603950,Russia}
\affiliation{Rzhanov Institute of Semiconductor Physics SB RAS, Novosibirsk 630090, Russia}
\author{Valerii\,M.\,Vinokur}
\affiliation{Materials Science Division, Argonne National Laboratory, Argonne, IL 60439, USA}
\author{Philip\,Kim}
\affiliation{Department of Physics, Harvard University,
    Cambridge, MA 02138, USA}
\begin{abstract}
We fabricate van der Waals heterostructure devices using few unit cell thick Bi$_2$Sr$_2$CaCu$_2$O$_{8+\delta}$ for magnetotransport measurements. The superconducting transition temperature and carrier density in atomically thin samples can be maintained to close to that of the bulk samples. As in the bulk sample, the sign of the Hall conductivity is found to be opposite to the normal state near the transition temperature but with a drastic enlargement of the region of Hall sign reversal in the temperature-magnetic field phase diagram as the thickness of samples decreases. Quantitative analysis of the Hall sign reversal based on the excess charge density in the vortex core and superconducting fluctuations suggests a renormalized superconducting gap in atomically thin samples at the 2-dimensional limit.
\end{abstract}

\maketitle

\label{Introduction}
Tunable van der Waals (vdW) structures enable the study of unconventional electronic properties of low-dimensional superconductivity (SC)~\cite{MagicAngleSuperconductor}. Measurement of the Hall effect, one of the most informative tools for probing electronic properties of low-dimensional systems, has renewed recent interest in the context of the particle-hole asymmetry and the Bose metal\,\cite{HallTaN, HallNbN, Breznay:2017, Phillips2018, Shahar2018} with the vanishing of Hall resistance $R_{xy}$ \cite{Breznay:2017}. One of the striking properties of SC is the sign change of the Hall resistance. As temperature $T$ decreases through the fluctuation region approaching the transition temperature $T_c$, the Hall resistivity decreases and changes its sign relative to the normal state. The Hall sign reversal in SC has been attributed to superconducting fluctuations (SF) for $T>T_c$ ~\cite{HallFinkelstein, HallTaN, HallNbN, HallMoN}) and vortex contributions for $T<T_c$ ~\cite{AoThoulDSC, KhomskiiDSC, PhysicaC:1994, JETPL:1995}. The Hall voltage exhibits a negative minimum and eventually reaches zero at low temperatures where vortices are completely immobilized\,\cite{ FisherDSC, FisherDorseyDSC}. 

The non-vanishing vortex contribution to the Hall signal is of special importance in high temperature superconductors (HTS), and is controlled by the magnitude of the superconducting gap $\Delta(T)$~\cite{KhomskiiDSC, PhysicaC:1994, JETPL:1995}. There have been striking observations that the superconducting gap (not pseudogap) $\Delta(T)$ is renormalized on approach to $T_c$ from below, obtained from angle-resolved photoemission spectroscopy (ARPES) of HTS and from tunneling spectroscopy of conventional low-$T_c$ films~\cite{NatComTiN}. Employing atomically thin vdW HTS with high crystallinity, we now can address the Hall sign reversal in the 2-dimensional (2D) limit, where fluctuation effects become significant. 

In this letter, we report fabrication of electronic devices based on atomically thin \chem{Bi_2Sr_2CaCu_2O_{8+\delta}} (BSCCO) SC samples and their magnetotransport properties in a wide temperature range. We observe enhanced fluctuation effects in these samples, manifesting as a large region of Hall sign reversal in the temperature-magnetic field phase diagram. We present quantitative description of the observed magnetotransport, considering SF above $T_c$ and vortex core contributions below $T_c$. Our analysis suggests that the renormalized superconducting gap remains finite at $T_c$.
\begin{figure}[H]
	\begin{center}
		\includegraphics[width=1\linewidth]{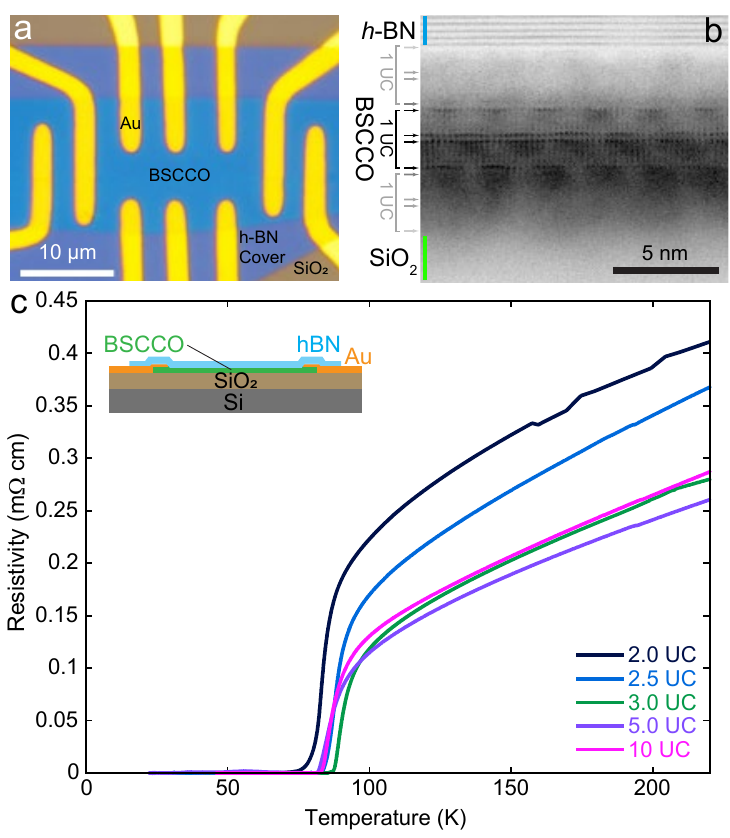}
		\setlength{\abovecaptionskip}{-10pt}
		\caption
		{\textbf{Van der Waals BSCCO device.}
			a. Optical image of Hall bar device, showing BSCCO with contacts and hexagonal boron nitride (\hbn) cover, as drawn in the inset below. b. Bright field scanning transmission electron microscopy image of cross-section of device. Columns of atoms are visible as dark spots. The layered structure of BSCCO and h-BN are visible, as is supermodulation of the BSCCO lattice. Black arrows point to location of bismuth oxide layers (darkest spots), while gray arrows show their expected positions. c. Resistivity as a function of temperature for vdW devices of different thicknesses.}
		\label{Fig1}
	\end{center}
\end{figure}

We prepare our few-unit-cell (UC) thick BSCCO using mechanical exfoliation in argon filled glovebox. The BSCCO system is technically challenging to handle in ambient condition, since BSCCO chemically interacts with water vapor in air \cite{Vasquez} and contains interstitial oxygen dopants which become mobile above 200K \cite{Poccia}. After conventional nano-fabrication steps, BSCCO typically becomes insulating \cite{Sandilands}. To address these issues we have developed a high-resolution stencil mask technique (See Supplementary Information), allowing us to fabricate samples in an argon environment without exposure to heat or chemicals, and subsequently sealed with a hexagonal boron nitride ($h$-BN) crystal on top. This technique solves the challenging problem of controlling the desired thickness~\cite{Bozovic2011} of BSCCO crystals, achieving a precision of 0.5 UC.  Fig.~\ref{Fig1}b shows a cross-sectional bright field scanning transmission electron microscope (STEM) image of a typical vdW heterostructure, where individual columns of atoms are clearly visible as dark spots. The darkest of these are bismuth atoms (arrows), which scatter the probing electrons most strongly. A supermodulation in the atoms is clearly visible, with a periodicity that agrees with the bulk value. Extrapolating from the position of the BiO layers, we see the outermost 1 UC on both sides become amorphous in this sample. This degradation of the top and bottom layers is likely to be present in all our samples, although its extent is likely sample-dependent. Fig.~\ref{Fig1}a shows a typical Hall bar.

Fig.\ref{Fig1}(c) shows the resistivity $\rho$ as a function of temperature $T$ for BSCCO devices between 2 - 10 UC. $\rho (T)$ exhibits a superconducting transition around 85 K, the measured bulk value prior to exfoliation, with a linear $T$-dependence in the normal region consistent with BSCCO near optimal doping~\cite{HTSbook}. At given temperature $T$, we find that $\rho$ increases as the thickness of the sample $d$ decreases, suggesting that thinner samples become poorer conductors. The surface degradation observed in the TEM image is presumably responsible for increasing $\rho$.

To quantitatively determine the SC transition temperature $T_c$ from $\rho(T)$, we adopt the SF framework for $T>T_c$~\cite{LarVar_book, TiN_3D2D, Bi-2212MT}. Here we take into account all three SF contributions: Aslamazov-Larkin, DOS and the dominant Maki-Thompson contributions \cite{Bi-2212MT, TiN_JETP}, using both $T_c$ and the pair-breaking parameter $\delta=h/16k_BT\tau_\phi$ as two fitting parameters. We assume the phase-breaking time to be $\tau_\phi \sim T^{-1}$~\cite{AA_review}. For all samples, the extracted $T_c$ is very close to the temperature where resistance falls fastest \cite{TiN_JETP, BenfattoTc}, and is consistent between samples. The values we obtain from this analysis for the samples with different $d$ are provided in the Supplementary Materials.

\begin{figure}[]
	\begin{center}
		\begin{center}
			\includegraphics[width=1.0\linewidth]{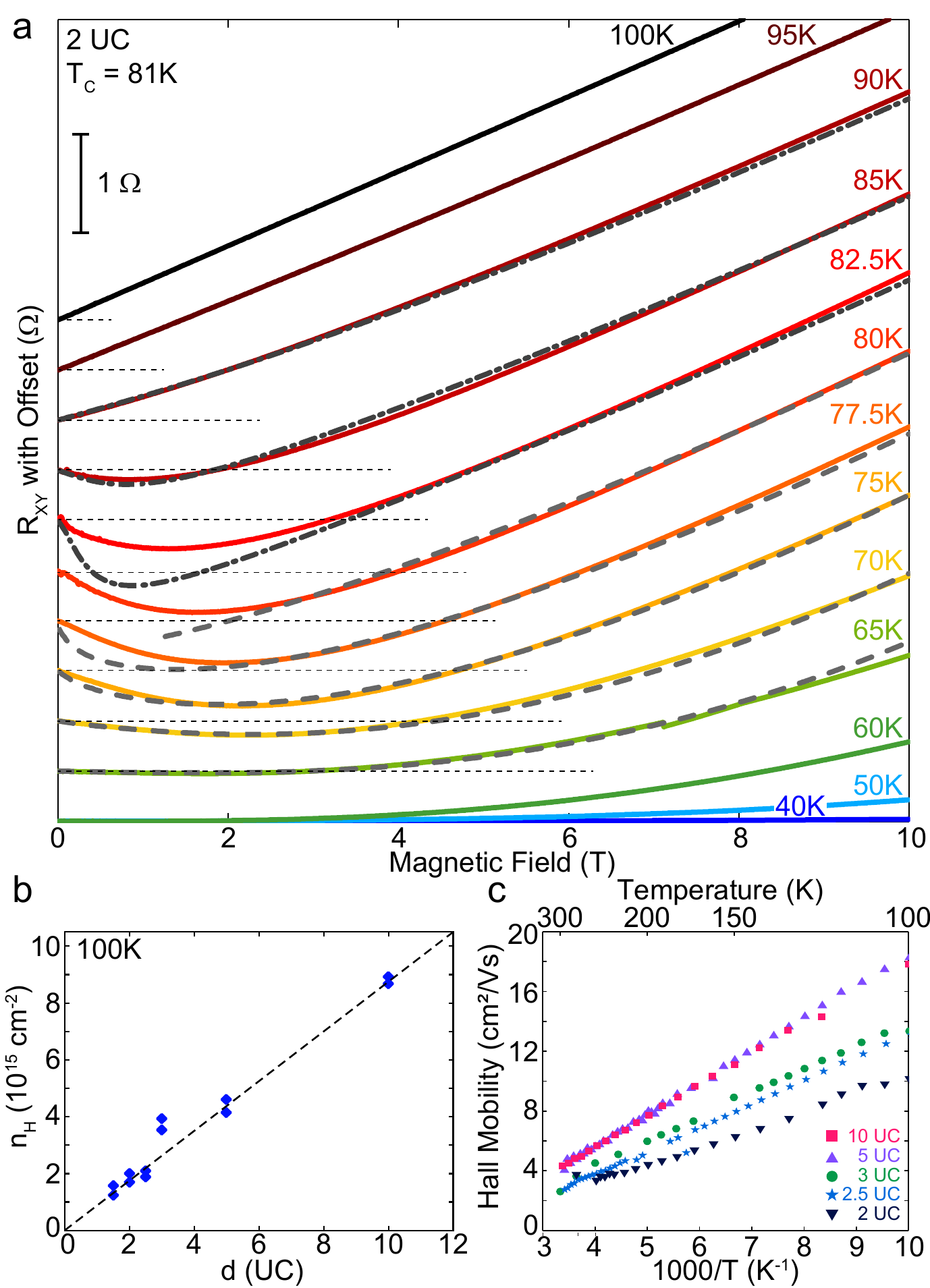}
		\end{center}
		\setlength{\belowcaptionskip}{-20pt}
		\caption
		{
			\textbf{Hall effect measurements} a. Hall resistance for a 2 UC sample. The curves are vertically shifted for clarity, with zeros indicated by dashed lines. Below 60K, the Hall effect has the same sign as in the normal state. Above 60K the sign reversal appears at magnetic fields $B<5$ T. b. Carrier density increases linearly with sample thickness in our devices, demonstrating good oxygen dopant retention down to 2 UCs. Data taken at 100K. c. Device mobility increases as samples become thicker, eventually saturating at 5 UC.
		}
		\label{Fig2}
	\end{center}
\end{figure}

Our ability to precisely control device thickness allows us to measure the Hall density $n_H$~\cite{Note}. Fig.~\ref{Fig2}(a) presents Hall data for a 2 UC device, where we took the odd component of $R_{xy}(B)$ to account for device geometric effects. In the normal state far above $T_c$ ($T \geq 100$ K), the Hall resistance $R_{xy}$ is linear in applied magnetic field $B$, allowing us to extract the Hall density $n_H=d/ecR_{xy}$. Fig.~\ref{Fig2}(b) shows $n_H$ measured at 100\,K, well above the transition temperature for samples with different $d$ shown in Fig. \ref{Fig1}(c). The Hall density $n_H$ scales linearly with $d$, demonstrating an excellent oxygen dopant retention in each CuO$_2$ plane, even in the degraded surface layers. The 3 UC sample deviates from this trend with more carriers than is expected, which agrees with the slightly increased $T_c$ compared to the others (see Fig.~\ref{Fig1}(c)). We also estimate the carrier Hall mobility $\mu_H=d/n_He\rho$ as shown in Fig. 2(c). Below 5 UC, $\mu_H$ decreases with $d$, indicating increasing disorder in thinner samples. We also notice that all samples empirically exhibit the trend $\mu_H \sim T^{-1}$ for $T \gg T_c$, suggesting that the normal carrier momentum relaxation time is $\tau_p \sim T^{-1}$ in our samples regardless of $d$.

As temperature decreases, the linear $R_{xy}(B)$ far above $T_c$ starts to develop a strong nonlinearity around $T_c$ (Fig. \ref{Fig2}(a)). Just above $T_c$, $R_{xy}(B)$ reverses sign at small $B$, reaching a minimum before increasing again with $B$. As $T$ continues to decrease, the dip in $R_{xy}(B)$ continues to broaden, reaching maximal size around 75 K, at which the sign reversal only vanishes by $B_0 = 4.7$ T. However, as $T$ decreases further, the Hall sign reversal weakens as both its magnitude and $B_0$ decrease. The regime of Hall sign reversal vanishes completely around 60~K, below which temperature $R_{xy}(B)$ remains positive for all magnetic fields, even as $R_{xy}$ decreases in magnitude and vanishes around 40~K.

\begin{figure}[]
	\begin{center}
		\begin{center}
			\includegraphics[width=1.0\linewidth]{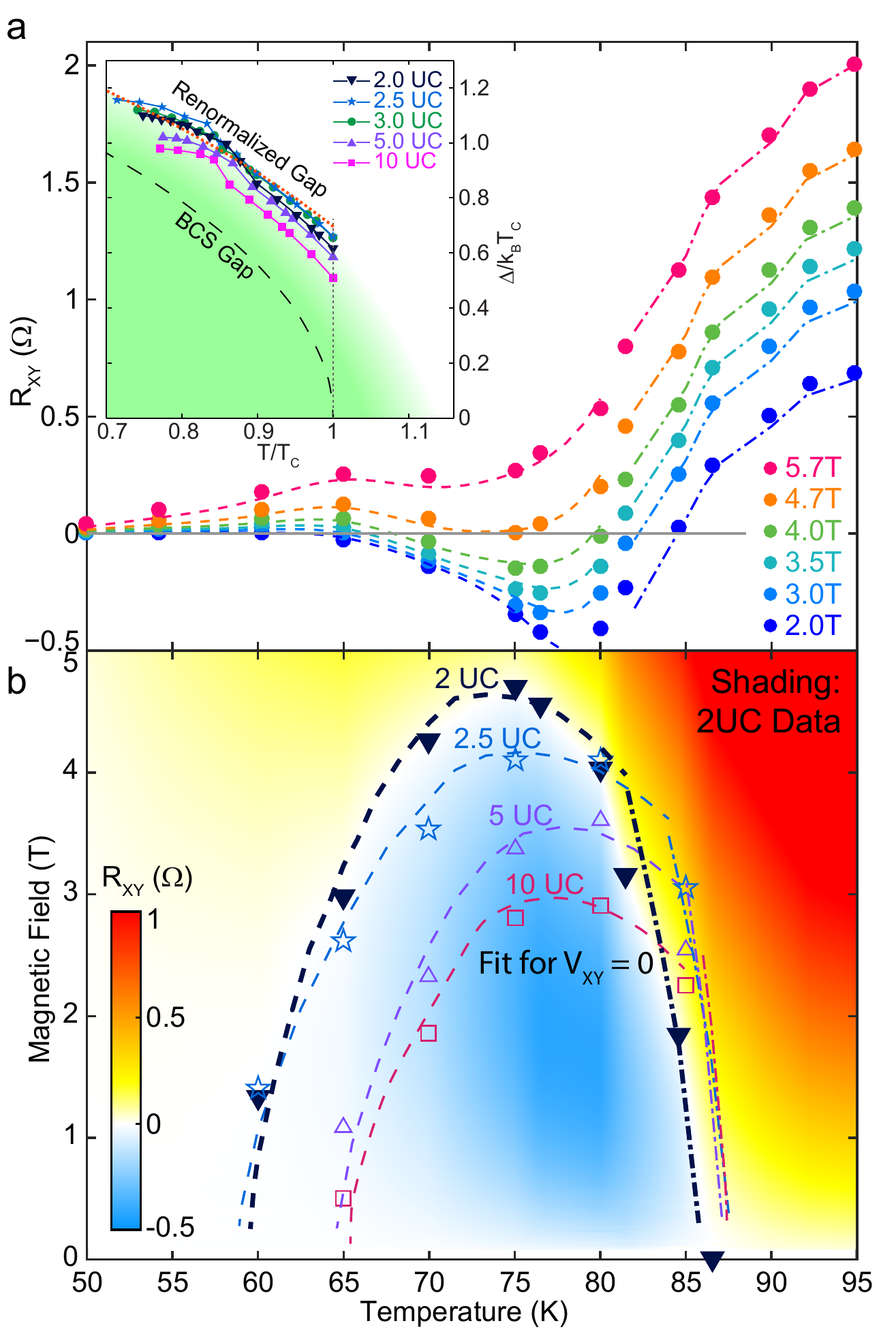}
		\end{center}
		\setlength{\belowcaptionskip}{-20pt}
		\caption
		{{\bf The double sign change}.
			a. $R_{xy}(T)$ at fixed magnetic fields for the 2UC device. Fits above (dash-dot) and below (dashed lines) $T_c$ are superimposed on experimental data (symbols).
			Inset: Superconducting gap extracted from fits below $T_c$ for all samples using Eq.\,(\ref{sigmaXY}).
			The lower dashed line is the BCS gap $\Delta(T) = 1.76 k_B T_c \sqrt{1-T/T_c}$
			with $T_c$ extracted from $R_{xx}(T, B=0)$.
			The renormalized gap curve is generated using the BCS equation, but with an elevated $T_{c0}$ to approximate the gap
			$\Delta(T)$ extracted from the fits. b. The Hall sign reversal phase diagram. Shading shows Hall resistance $R_{xy}(B,T)$ for a 2UC device. Blue region indicates the area of negative Hall resistance. Symbols show the locus $R_{xy}=0$ for different sample thicknesses, with the dashed (dash-dot) lines generated from fits below (above) $T_c$ (see SI). As device thickness decreases, the Hall-sign-reversed region becomes larger.
		}
		\label{Fig3}
	\end{center}
\end{figure}


Fig. \ref{Fig3}(a) shows the evolution of $R_{xy}(T)$ at constant $B$, highlighting a double sign reversal in $R_{xy}$. For instance, at $B \approx 4$~T, it is clear that $R_{xy}(T)$ changes sign twice as $T$ increases, once around $T \approx 67$~K and again at 80 K. The complete phase diagram for the Hall sign reversal is shown in Fig. \ref{Fig3}(b), where we have superimposed the Hall reversal boundary for different samples with different $d$, yet very similar $T_c$. The region where we observe the negative Hall effect is a well-defined domain of the $T-B$ phase diagram. The region of $R_{xy}(T, B)<0$ grows noticeably as $d$ decreases, indicating that fluctuations enhance Hall sign reversal.

The region of the Hall sign reversal for few-unit-cell BSCCO is distinctly different from that of bulk samples. In bulk HTS samples, Hall sign reversal is observed only in the vortex liquid domain, i.e. in the strip between the vortex lattice melting line $B_m(T)$ and the $H_{c2}(T)$ line. Near $B_m(T)$, the Hall resistance is exponentially suppressed, and the Hall sign reversal region often completely lies within $T<T_c$ ~\cite{DoubleSignFit:1998}. On the other hand, in conventional superconductors, $B_m(T)$ and the $H_{c2}(T)$ practically coincide; thus usually all Hall sign change is assigned to the fluctuation region. In our atomically thin BSCCO, unlike the bulk samples, the Hall sign reversal region occurs across $T_c$. Moreover, we observe no sudden changes in $R_{xy}(T)$ upon crossing $T_c$, and the region of Hall effect sign reversal falls both above and below $T_c$ in all samples (Fig.~\ref{Fig3}b). This calls for a universal approach to the description of 2D superconductivity, which can be formulated in the framework of the Keldysh technique\,\cite{AK}. Above $T_c$, in the fluctuation regime, this approach simplifies to the quantum kinetic equation\,\cite{HallFinkelstein}, where the quantum corrections to conductivity are provided by the Gaussian approximation~\cite{HallFinkelstein, HallNbN}. For $T<T_c$, the Keldysh action can be reduced to the phenomenological form explicitly accounting for the vortex excitations and normal carriers' contributions\,\cite{PhysicaC:1994, JETPL:1995}.

Qualitatively, superconducting fluctuations are Cooper pair fluctuations with a finite lifetime, arising above \Tc. Under applied magnetic field, these pairs rotate around their center of mass \cite{EPL_rotatingFCP} and can be viewed as elemental current loops \cite{VoticesAboveTcNature, VoticesAboveTcPRB}. Applied external current exerts Magnus force moving these loops along circular paths. This gives rise to a Hall voltage opposite to that from the normal carriers. More quantitatively, the SF contribution to Hall conductivity manifests as a negative correction $\delta \sigma_{xy}$ to the positive normal component $\sigma_{xy}^n$ \cite{LarVar_book, HallFinkelstein}: $\sigma_{xy} = \sigma_{xy}^n + \delta \sigma_{xy}$. Within the Drude framework, $\sigma_{xy}^n $ can be estimated from experimentally accessed quantity $\sigma_{xy}^n \approx \frac{en_H\mu_H^2}{c}B$. Quantitative expression for $\delta\sigma_{xy}$ can be expressed using the Gaussian approximation~\cite{HallFinkelstein}:
\begin{equation}
\delta\sigma_{xy}=\frac{2e^2k_BT}{hd}\zeta f(D,B,T) \label{dSigma}
\end{equation}
where $D$ is the normal carrier diffusion coefficient, $f$ is a dimensionless function whose explicit form is given in the Supplementary Information, and $\zeta$ is a parameter accounting for particle-hole asymmetry in the time-dependent Ginzburg-Landau equation. $\zeta$ is expressed as the change of $T_c$ with respect to the chemical potential $\mu$: $\zeta = -\frac{1}{2} \partial (\ln T_c)/\partial \mu \approx 1/(\gamma E_F)$~\cite{Varlamov:1999, LarVar_book, HallFinkelstein}. Here $\gamma$ is the dimensionless coupling constant parameterizing the attractive electron-electron interaction that induces superconductivity.

As temperature decreases, the SF contribution $\delta\sigma_{xy}$ increases, leading to deviation from linear $R_{xy}$\,vs.\,$B$ behavior, and eventually to the sign change of $\sigma_{xy}$ as soon as $\delta\sigma_{xy}$ starts to dominate\,\cite{HallTaN, HallNbN, HallMoN}. In a diffusive metal, $D \approx \frac{2}{3}\mu_H E_F$. Thus, Eq. (\ref{dSigma}) can be fit into the experimentally measured $\sigma_{xy}$ using $E_F$ and $\gamma$ as two fitting parameters in a wide range of $B$ and $T>T_c$. For this analysis, we also employ previously measured $\mu_H$. As shown in Fig.~\ref{Fig2}(a) (dash-dot lines), this model fits our data very well above $T_c$. The numerical values of our fitting parameters $E_F$ and $\gamma$ are summarized in Table I in Supplementary Information. We obtain $E_F \approx 0.5$ eV. This is in reasonable agreement with the literature value, considering the fact that $E_F$ of cuprates is often an order of magnitude larger than the superconducting gap $\Delta(0)$ \cite{HTSC_EF} and $E_F\sim 0.1$ eV for \chem{La_{2-x}SrCuO_2}~\cite{HTSC_EF}. The value $\gamma \approx 0.1$ corresponds to the weak coupling limit.

We now turn our attention to the Hall sign reversal in the temperature range $T<T_c$. The challenge of describing the Hall effect below $T_c$ is in producing a thorough account of all the contributions to vortex dynamics. A comprehensive description of the Hall conductivity $\sigma_{xy}$ explicitly including topological aspects of vortex dynamics (Berry phase), normal carrier scattering, and weak pinning effect was developed in\,\cite{PhysicaC:1994, JETPL:1995}, where Hall conductivity acquires the form:
\begin{equation}
\sigma_{xy}=  \frac{ \Delta^2 \cdot n_0\cdot ec}{E_F^2 \cdot B} [(
\tau \Delta /\hbar)^2g-\textrm{sign}(\delta
n)]+\sigma^n_{xy}(1-g), \label{sigmaXY}
\end{equation}
where $n_0$ and $n_{\infty}$ are the normal carrier density inside and outside the vortex core respectively, and $\delta n = n_0-n_{\infty}$ is the excess charge inside the vortex; $\tau$ is the relaxation time of the normal carrier in the vortex core; and parameter $g$ expresses the superconducting fraction of the carriers. In this work, we consider a two-fluid model of a $d$-wave symmetry superconductor \cite{Tinkham} so that $g(T) = 1-(T/T_c)^2$.

The physical origin of the Hall effect sign change in this low-temperature regime is due to the excess charge $\delta n$ of the vortex core~\cite{KhomskiiDSC, PhysicaC:1994}. The difference in carrier density $\delta n/n_0$ is of the order of $(\Delta/E_F)^2$~\cite{PhysicaC:1994, JETPL:1995, DoubleSignFit:1998}. Here, the sign of the vortex contribution to the Hall effect is determined by the relation between sign($\delta n$) and  $ \tau \Delta$. Since the Hall sign is reversed in the regime $T<T_c$, this observation empirically fixes $\textrm{sign} (\delta n)=1$. Then, the first term in Eq. (\ref{sigmaXY}), the vortex core contribution $\sigma^{vc}_{xy}$, can be negative as $ \Delta(T) < \hbar/\tau$. From this definition, we also note that $\sigma_{xy}\sim B^{-1}$ while $\sigma^n_{xy} \sim B$. Therefore, the total Hall sign reversal is expected at low magnetic fields, where negative vortex contribution $\sigma^{vc}_{xy}$ dominates the positive normal carrier contribution $\sigma_{xy}^n$.

We can compare Eq.~(\ref{sigmaXY}) with our experimental data for $T<T_c$ quantitatively. In order to fit experimental curves with Eq.~(\ref{sigmaXY}), we estimate the normal contribution $\sigma_{xy}^n$ below $T_c$ using our empirical observation that $\mu_H \sim T^{-1}$ in the normal state above $T_c$. Extrapolating this relation to $T<T_c$ in the two-fluid picture, we assume $\sigma_{xy}^n(T) = \sigma^n_{xy}(T_0)(T_0/T)^2$, with $T_0=100$ K for our analysis. Then, Eq. (\ref{sigmaXY}) can be used to fit our data shown in Fig.~\ref{Fig2}(a) and Fig.~\ref{Fig3}(a) (for fixed $T < T_c$ and $B$ respectively), using $\tau$, $n_0$ and $\Delta(T)$ as fitting parameters. The values $E_F$ and $T_c$ were previously determined from analysis of $R_{xy}(B,T)$ at $T>T_c$ with SF theory. The parameter $n_0 \approx 10^{21}$ cm$^{-3}$ agrees with the widely accepted value for the cuprates \cite{Bozovic_n0}. The relaxation rate of the normal carriers in the vortex core is estimated to be $\tau \approx 0.1$ ps. This value is in reasonable agreement with the quasiparticle lifetime estimated from the scanning tunneling spectroscopy of the vortex cores in BSCCO \cite{tau_vortex}, where normal quasiparticle excitations at $E \approx 7$~meV was reported. A crude estimate of the core state lifetime is therefore $\hbar/E \approx$ 0.1 ps. The numerical values of all our fitting parameters are summarized in Table I in Supplementary Information.

Dashed lines in Fig.~\ref{Fig2}(a) and Fig.~\ref{Fig3}(a) are fitted lines calculated according to Eq.~(\ref{sigmaXY}). Here, importantly, we kept the temperature dependence of the superconducting gap $\Delta(T)$ as a free fitting parameter. This was prompted by two reasons. First, setting the classic BCS value of $\Delta(T/T_c)$ in Eqs.~(\ref{sigmaXY}), we would obtain unreasonably small values of the field $B$ where the sign reversal occurs. Second, the fact that the \textit{superconducting} gap (not pseudogap) $\Delta(T)$ is nonzero at $T_c$ is theoretically proposed \cite{PRLGap, NatComTiN} and experimentally observed in tunneling~\cite{SciRepGap} and in angle-resolved photoemission spectroscopy (ARPES) ~\cite{NatPhysGap}. Our estimated temperature dependences of superconducting gap $\Delta(T/T_c)/T_c$ are shown in inset of Fig.~\ref{Fig3}a for all samples. We notice that $\Delta(T/T_c)$ dependence differs from the standard BCS dependence, namely $\Delta(T_c)\neq 0$. The deviation from BCS is more pronounced for thinner samples, suggesting that the fluctuation effects may be the major source of such large deviation. Phenomenologically, it is interesting to note that the estimated $\Delta(T)$ evolves according to the expected BCS equation, but with a $T_{c0}$ temperature which is about 10 percent larger than the observed $T_c$, suggesting renormalization of the SC gap.

Finally, using the same set of fitting parameters, we can identify the phase boundary of the Hall sign reversed region in Fig.~3(b) for further independent comparison with experiment. The sign reversal locus, $R_{xy} =0$, according to Eq.\,(\ref{sigmaXY}) is defined by the relation:
\begin{equation}
B^2 = \left(\frac{\Delta}{E_F}\right)^2\frac{n_0c}{n_H\mu_H^2}\frac{[(\Delta\tau/\hbar)^2g-1]}{1-g}
\label{Field}
\end{equation}
The region defined by Eq.\,(\ref{Field}) demonstrates excellent agreement with the experimental observation shown in Fig. \ref{Fig3}(b) for $T<T_c$. Above $T_c$, however, the phase boundary drops rapidly as $T$ increases, a fact accurately captured in our SF fits.

In conclusion, we developed van der Waals heterostructure assembly techniques specialized to the cuprates. We fabricated few-unit-cell \chem{Bi_2Sr_2CaCu_2O_{8+\delta}} crystals, where strongly enhanced Hall sign reversal was observed. From quantitative analysis of the double Hall sign reversal, we find that the superconducting gap is nonzero at the critical temperature $T_c$.

The experiments at Harvard was supported by National Science Foundation (DMR-1809188) and the Gordon and Betty Moore Foundation EPiQS Initiative (GBMF4543). Stencil masks were fabricated at the Harvard Center for Nanoscale Systems (CNS), a part of NNCI, NSF award 1541959. S.Y.F.Z. was partially supported by the NSERC PGS program. NP was partially supported by ARO (W911NF-17-1-0574). G.D.G. is supported by the Office of Science, U.S. Department of Energy under Contract No. de-sc0012704. R.Z. is supported by the Center for Emergent Superconductivity, an Energy Frontier Research Center funded by the U.S. Department of Energy, Office of Science. K.W. and T.T. acknowledge support from the Elemental Strategy Initiative conducted by the MEXT, Japan and the CREST (JPMJCR15F3), JST. The work of V.M.V. was supported by the US Department of Energy, Office of Science, Basic Energy Sciences, Materials Sciences and Engineering Division. The work of S.V.P. on the analysis of the experimental data was supported by the Russian Science Foundation under Grant No. 15-12-10020.


\end{document}